\documentclass[aps, 
twocolumn,review,
groupedaddress ,bibnotes]{revtex4-2}

\usepackage{mathrsfs}
\usepackage{amssymb}
\usepackage{amsmath}
\usepackage{amsfonts,bm}
\usepackage[  colorlinks=true,
    linkcolor=blue,
    citecolor=blue]{hyperref}

\usepackage{pgffor}
\foreach \x in {A,...,Z}{
  \expandafter\xdef\csname  b\x\endcsname{\noexpand\mathbb{\x}}
}
\foreach \x in {A,...,Z}{
  \expandafter\xdef\csname c\x\endcsname{\noexpand\mathcal{\x}}
}
\foreach \x in {A,...,Z}{
  \expandafter\xdef\csname s\x\endcsname{\noexpand\mathscr{\x}}
}
\foreach \x in {A,...,Z}{
  \expandafter\xdef\csname sf\x\endcsname{\noexpand\mathsf{\x}}
}
\foreach \x in {a,...,z}{
  \expandafter\xdef\csname sf\x\endcsname{\noexpand\mathsf{\x}}
}
\foreach \x in {A,...,Z}{
  \expandafter\xdef\csname  fk\x\endcsname{\noexpand\frak{\x}}
}
\foreach \x in {a,...,z}{
  \expandafter\xdef\csname  fk\x\endcsname{\noexpand\frak{\x}}
}

  \newcommand{\Tr}{\text{Tr}}
  
  \newcommand{\EV}[1]{\langle #1 \rangle}
  
    \newcommand{\EB}[1]{( #1 )}
  
  \newcommand{\ket}[1]{\ensuremath{\left|#1\right\rangle}}
  \newcommand{\bra}[1]{\ensuremath{\left\langle#1\right|}}

\newcommand{\be}{\begin{equation}}
\newcommand{\ee}{\end{equation}}
\newcommand{\bpm}{\begin{pmatrix}}
\newcommand{\epm}{\end{pmatrix}}

\newcommand{\beqn}{\begin{eqnarray}}
\newcommand{\eeqn}{\end{eqnarray}}

\begin{document}

\title{Macdonald Index from VOA and Graded Unitarity}

\author{Hongliang Jiang}
 \email[]{jianghongliang@fudan.edu.cn}
\affiliation{Center for Mathematics and Interdisciplinary Sciences, Fudan University, Shanghai 200433,
China}
\affiliation{Shanghai Institute for Mathematics and Interdisciplinary Sciences (SIMIS), Shanghai 200433,
China}

\begin{abstract}

The SCFT/VOA correspondence provides a powerful framework for studying 4d $\mathcal N=2$ superconformal field theories (SCFTs) through the mathematical machinery of 2d vertex operator algebras (VOAs). It captures the Schur operators of the underlying SCFT, whose spectrum is encoded by the Schur index and its refinement, the Macdonald index. While the Schur index is identified with the vacuum character of the associated VOA, a general VOA-based derivation of the Macdonald index has remained elusive.
In this letter, we propose a novel and intrinsic method for recovering a special non-Schur limit of the Macdonald index directly from the VOA. The construction requires no additional assumptions and applies whenever the underlying 4d theory is unitary. We test the proposal in a variety of examples, and further extend it to the case with surface defects, suggesting a notion of graded unitarity in the presence of defects. Our method also introduces a new class of series for general VOAs, analogous to but distinct from the conventional character,  and potentially useful in broader contexts.

\end{abstract}

\maketitle

\section{Introduction}

Superconformal field theories play a central role in modern theoretical physics and mathematics. They not only provide a powerful window into strongly coupled quantum field theories and quantum gravity through the AdS/CFT correspondence, but also encode rich mathematical structures of independent interest. One celebrated example is the SCFT/VOA correspondence \cite{Beem:2013sza}, which associates to every 4d \(\cN=2\) superconformal field theory (SCFT) a 2d vertex operator algebra (VOA). This correspondence offers a powerful framework for studying SCFTs by importing a wide range of techniques from mathematics. For example, the stress tensor and flavor current in 4d are mapped to the stress tensor and current in 2d, which are governed by the well-studied Virasoro and Kac--Moody algebras.

However, the SCFT/VOA correspondence presents at least two important puzzles and open questions. The first is a conceptual puzzle concerning unitarity. The central charge and level in the 4d SCFT and the 2d VOA are related by
\be\label{ack24d}
c_{2d}=-12c_{4d}, \qquad k_{2d}=-\frac12 k_{4d}.
\ee
The minus signs in these relations are puzzling, since they imply that a unitary 4d SCFT is mapped to a 2d VOA that is non-unitary in the conventional sense. This apparent loss of unitarity, which is a powerful constraint in both physics and mathematics, makes the correspondence conceptually obscure.

The second question is more practical in nature. For a 4d \(\cN=2\) SCFT, one can define the Schur index \cite{Gadde:2011ik}
\be\label{IdS}
\cI_S(q)=\cI_M(q,1)=\Tr_{\cH_{\mathrm{Schur}}} (-1)^F q^{h},
\ee
where the trace is taken over the full space of Schur operators, namely the subset of protected operators whose quantum numbers satisfy
\be\label{Schcond}
E=2R+j_1+j_2, \qquad r=j_2-j_1.
\ee
Here \(E\) is the conformal dimension; \(R\in \frac12 \bN\) and \(r\in \frac12 \bZ\) are the charges under \(SU(2)_R\) and \(U(1)_r\), respectively; and \(j_1,j_2\in \frac12 \bN\) are the Lorentz spins under \(\fks\fku(2)\times \fks\fku(2)\simeq \fks\fko(4)\). We also define \(h\equiv E-R\in \frac12\bN\), which is identified with the conformal weight in the associated 2d VOA.

According to the SCFT/VOA correspondence, the Schur index of the 4d theory is identified with the vacuum character of the VOA. However, one may also define the Macdonald index \cite{Gadde:2011uv},
\be
\cI_M(q,T)=\Tr_{\cH_{\mathrm{Schur}}} (-1)^F q^{h} T^{R-r},
\ee
which may be regarded as a refinement of the Schur index. 
This naturally leads to the second question: is there an intrinsic way to compute the Macdonald index purely within the framework of the VOA, just as for the Schur index? This is a natural question, since the Macdonald index counts exactly the same set of operators as the Schur index. The difficulty, however, is that the Macdonald index depends on the \(SU(2)_R\) quantum number of Schur operators, and this information is obscured in the VOA description due to the topological twist. Although several proposals for recovering the Macdonald index from the VOA have been put forward \cite{Song:2016yfd,Bhargava:2023hsc,Andrews:2025krn,Kang:2025zub}, their applicability is limited and they rely on assumptions that are not   justified.

These two issues have remained open for more than a decade since the proposal of the SCFT/VOA correspondence. Very recently, a new notion of graded unitarity was introduced in \cite{ArabiArdehali:2025fad,Beem:2026lkq}, providing a natural explanation for the puzzling minus signs in Eq.~\eqref{ack24d}. In this letter, we make progress on the second question. More precisely, we propose a method to recover the Macdonald index purely from the VOA in a particular limit distinct from the Schur limit:
\be\label{IRind}
\cI_R(q)= \cI_M(qe^{\pi i},e^{\pi i})
=\Tr_{\cH_{\mathrm{Schur}}} (-1)^F q^{h} (-1)^{R-r+h  }.
\ee
Here the subscript \(R\) emphasizes that this index is closely tied to the \(SU(2)_R\) symmetry.

The two indices \eqref{IdS} and \eqref{IRind} can be viewed as special limits of the family of indices \(\cI_M(qe^{\pi i\alpha},e^{\pi i\alpha})\), at \(\alpha=0\) and \(\alpha=\pm 1\), respectively. This also provides a prescription for defining \(\cI_R\) when \(\cI_M\) involves fractional powers of \(q\) and \(T\). A nice feature of \(\cI_R(q)\) is that it is a series in \(q\) or \(q^{\frac12}\) with integer coefficients, thanks to the relation
$
R-r+h=2(R+j_1)\in \bZ,
$
which follows from \eqref{Schcond}.

Our central result is an intrinsic method for computing \(\cI_R\) entirely within the framework of the VOA, without any additional assumptions. It therefore applies universally to all unitary 4d  $\mathcal N=2$ SCFTs. More broadly, our proposal points to a genuinely new type of quantity in VOAs, similar to but distinct from the traditional character. This object is of independent interest, may find further applications in both physics and mathematics, and deserves further investigation.

In Section~\ref{prop}, we present our main proposal for computing the  special limit of the Macdonald index intrinsically from the VOA. In Section~\ref{exg}  we test the proposal in a range of explicit examples, and then generalize it to the case with defects in Section~\ref{defect}. Finally, we conclude in Section~\ref{conl}. Some technical details are deferred to the appendices, while further details and generalizations will be presented in future work.

\bigskip

\section{\label{prop}Proposal }   \label{prop}
According to the SCFT/VOA correspondence, the Schur operators of an SCFT are organized into a VOA, \(\cV\), whose structure is encoded in the operator product expansion (OPE):
\be\label{ABope}
A(z)B(0)\sim \sum_{n=1}^{h_A+h_B}\frac{\{AB\}_n(0)}{z^n}.
\ee
The OPE should be associative. In actual    OPE computations, we will make extensive use of the package $ \mathtt{OPEdefs}$~\cite{Thielemans:1991uw}.
Note that the most singular term in the OPE, corresponding to \(n=h_A+h_B\), is either zero or proportional to the identity operator. By a slight abuse of notation, we will also use \(\{AB\}_{h_A+h_B}\) to denote the coefficient after stripping off the identity operator. The operators in \(\cV\) carry weight \(h\), \(U(1)_r\) charge \(r\), fermion parity \(f=0,1\), and possibly additional flavor charges, all of which are preserved by the OPE. The \(SU(2)_R\) charge, however, is not respected by the OPE, which is the main obstacle to recovering the Macdonald index from the VOA.

We would like to define a suitable inner product on the space of operators in the VOA. To do so, we introduce an anti-linear automorphism of the VOA \cite{dong2013unitaryvertexoperatoralgebras}. The anti-linear automorphism
\(\phi:\cV\to\cV\)
is an anti-linear map that leaves the unit operator and stress tensor invariant,
\(\phi(\mathbf{1})=\mathbf{1}\) and \(\phi(T)=T\). It satisfies
$
\phi(uA+vB)=u^*\phi(A)+v^*\phi(B),
$
and acts as an automorphism  
\be\label{phiAB}
\phi\big(\{AB\}_n\big)=(-1)^{f_Af_B}\{\phi(A)\phi(B)\}_n,
\ee
where the overall sign accounts for the fermion parity of the operators. Note that the action of \(\phi\) on \(\cV\) is completely determined by its action on the strong generators of \(\cV\). We further require that
\be\label{phisq}
\phi^2=(-1)^{2h}.
\ee
Moreover, \(\phi\) should reverse the \(U(1)_r\) charge of operators while preserving the conformal weight and fermion parity.

As we will see in the examples below, once the OPEs of the VOA are specified, these conditions—possibly together with certain \(R\)-charge selection rules—impose strong constraints on  \(\phi\).

The anti-linear automorphism allows us to define an inner product on the space of operators in \(\cV\):
\be\label{innpd}
\EB{A,B}= \{\phi(A)B\}_{h_A+h_B}~.
\ee
Under complex conjugation, one finds
\beqn
\EB{A,B}^*
&=&
\phi\!\left(\{\phi(A)B\}_{h_A+h_B}\right)
\\
&=&
 (-1)^{f_Af_B}\{\phi^2(A)\phi(B)\}_{h_A+h_B}
\\
&=&
\phi_A^2\,(-1)^{f_Af_B}\{A\phi(B)\}_{h_A+h_B}
\\
&=&
(-1)^{h_A+h_B}\phi_A^2\,\EB{B,A}~,
\eeqn
where we used   \eqref{phiAB} and  \eqref{phisq}, together with the property of the OPE \eqref{ABope}. 

In practice, we will only consider the inner product between two operators of the same weight, \(h_A=h_B\). In that case, we have
\be
\EB{A,B}^*=\EB{B,A},
\ee
so \(\EB{-,-}\) defines a Hermitian form.

If the underlying 4d theory is unitary, the physical operators should satisfy
\be\label{posO}
(\cO,\cO)\propto (-1)^{R-r+h},
\ee
where \(\propto\) denotes equality up to a positive multiplicative factor. We will test this statement in a variety of examples below. Its validity follows from the unitarity of the underlying 4d quantum field theory—or, equivalently, from reflection positivity in Euclidean signature—which is reformulated in the VOA language as graded unitarity following \cite{ArabiArdehali:2025fad}. A proof of \eqref{posO} is provided in Appendix~\ref{posOpf}.

This positivity condition \eqref{posO} allows us to compute the special limit of the Macdonald index introduced in \eqref{IRind}. The procedure is as follows:

\begin{enumerate}
\item For each fixed     $h,f $, enumerate all operators of weight \(h\) and fermion parity \(f\), denoted by \(\cO_i^{h,f}\).

\item Compute the inner products \(\EB{\cO_i^{h,f},\cO_j^{h,f}}\) and organize them into a Gram matrix \(M_{h,f}\).

\item Compute the eigenvalues of \(M_{h,f}\), and denote by \(x_{h,f}^+\), \(x_{h,f}^-\), and \(x_{h,f}^0\) the numbers of positive, negative, and zero eigenvalues, respectively.

\item Use these data to compute
\beqn\label{Ids}
I_S(q)&=&\sum_{h,f}(-1)^f q^h \big(x_{h,f}^+ + x_{h,f}^- \big),\\ \qquad
\label{Idr}
I_R(q)&=&\sum_{h,f}(-1)^f q^h \big(x_{h,f}^+ - x_{h,f}^- \big).
\eeqn
\end{enumerate}

Let us make a few comments. First, the inner product \(\EB{-,-}\) defines a Hermitian form, so the Gram matrix \(M_{h,f}\) is Hermitian and therefore has real eigenvalues. Second, the zero eigenvalues correspond to null operators with \((\cO,\cO)=0\), which are therefore not physical and should not contribute to the index. Third, the numbers of positive, negative, and zero eigenvalues of a Hermitian matrix are invariant under unitary transformations, and hence are independent of the choice of operator basis. In particular, one may choose a basis in which the Gram matrix is diagonal. The diagonal entries then encode precisely the sign information in \eqref{posO}, which is exactly the factor that appears in \eqref{IRind}.

The series \eqref{Ids} is therefore just the VOA character, up to an overall normalization, and coincides with the Schur index \eqref{IdS}. We will refer to the other series \eqref{Idr} as the modified character. Our central claim is that it reproduces the special non-Schur limit of the Macdonald index in \eqref{IRind}.

 \bigskip

\section{Examples}\label{exg}

We now verify our proposal in some examples, including free theories and various kinds of Argyres-Douglas  (AD) theories.

\subsection{Free vector}
We start with the free vector, whose VOA is given by the symplectic fermion, with OPE 
\[
\eta_+(z)\eta_-(0)\sim \frac{-1}{z^2} ,
\qquad
\eta_-(z)\eta_+(0)\sim \frac{ 1}{z^2} .
\]
 The   anti-linear automorphism can be found by imposing the defining conditions, and is given by 
\be \label{phivec}
\phi(\eta_+)=  t\eta_-,
\quad
\phi(\eta_-)= \frac{1}{t} \eta_+, \quad t\in \bR.
\ee
 The inner product can also be computed  using the definition  \eqref{innpd}:
\be\label{etaprod}
\EB{\eta_+,\eta_+}=t, \quad \EB{\eta_-,\eta_-}=-1/t , \quad \EB{\eta_+,\eta_-}=0.
\ee
Since we know that \(\eta_+\) and \(\eta_-\) arise from the multiplets \(\bar{\mathcal D}_{0(0,0)}\) and \(\mathcal D_{0(0,0)}\), respectively \cite{Dolan:2002zh}, they should have \(h=1\), \(R=\frac12\), and \(r=-\frac12,+\frac12\), respectively. Imposing \eqref{posO}, we find that \(t>0\). The precise value of \(t\) is not important, since it simply corresponds to the normalization of the operators  \footnote{Indeed, defining \(\underline{\eta_+}=\frac{1}{\sqrt{t}}\,\eta_+\) and \(\underline{\eta_-}=\sqrt{t}\,\eta_-\), we see that the OPEs of \(\underline{\eta_+}\) and \(\underline{\eta_-}\) are exactly the same as those of \(\eta_+\) and \(\eta_-\). Moreover, \(\phi(\underline{\eta_+})=\underline{\eta_-}\) and \(\phi(\underline{\eta_-})=\underline{\eta_+}\).}.

%
%

It is easy to see that   we can choose the following  basis for the operators:
\be
O_{\bm m, \bm n}=\prod_i(\partial^i\eta_+ )^{m_i}\prod_j(\partial^j\eta_-  )^{n_j}, \quad
 m_i,n_j=0,1.
\ee
The Gram matrix is diagonal with entries
\be\label{vecnom}
\EB{O_{\bm m, \bm n},O_{\bm m, \bm n}}\propto  (-1)^{\sum_j n_j +j n_j+\sum_i i m_i}.
\ee
The fermion parity is $f_{O_{\bm m, \bm n}}=(-1)^{\sum_i m_i+\sum_j n_j}$.

Following our proposal, we then find that 
 \beqn
 I_S&=&\prod_{i=1}^\infty (1-q^i)\prod_{j=1}^\infty (1-q^j)=(q,q)_\infty (q,q)_\infty,
 \\
\label{IRvec}
  I_R&=&\prod_{i=1}^\infty (1-(-1)^iq^i)\prod_{j=1}^\infty (1+(-1)^jq^j)=(q,q)_\infty (-q,q)_\infty,
  \qquad  \;
 \eeqn
which   indeed agree with the limits of the Macdonald index: 
\beqn
\cI_M&=&(q;q)_\infty\,(qT;q)_\infty .
\eeqn
It is worth emphasizing that the result \eqref{IRvec} is independent of the sign of \(t\) in \eqref{phivec}, simply because the pair \(\eta_+\) and \(\eta_-\), which have opposite inner products \eqref{etaprod}, both appear.

\subsection{Free hyper} 

The chiral algebra for the free hypermultiplet is given by symplectic boson   \cite{Beem:2013sza}, with OPE
\begin{equation}
q(z)\tilde q(0) \sim \frac{-1}{z}, \qquad
 \tilde q(z)  q(0) \sim \frac{ 1}{z}.
\end{equation}

We find that the anti-linear automorphism $\phi$ acts as
\be
\phi(q) =- t\tilde q, \quad \phi(\tilde q) =\frac{1}{t}q, \quad t\in \bR.
\ee
In particular, $\phi^2$ acts as $-1$ on $q,\tilde q$, in agreement with \eqref{phisq} because $h=\frac12$.

The non-zero inner products are given by
\be
\EB{q,q}=-t, \qquad
\EB{\tilde q, \tilde q}=-1/t.
\ee
Since we know that $q,\tilde q $  come from the  multiplet $\hat \cB_{\frac12} $ \cite{Dolan:2002zh}, they should have $h= R=\frac12$ and $r=0$.
Imposing  \eqref{posO}, we find that $t>0$.

The space of operators is spanned by 
\be
O_{\bm m, \bm n}=\prod_i(\partial^i q )^{m_i}\prod_j(\partial^j\tilde q )^{n_j}, \qquad
 m_i,n_j\in \mathbb N.
\ee
The Gram matrix is again diagonal:
\be
\EB{O_{\bm m, \bm n},O_{\bm m, \bm n}} \propto (-1)^{\sum_i i m_i +m_i+\sum_j j n_j+n_j}.
\ee
This then allows us to compute:
 \beqn
 I_S&=&\prod_{i=0}^\infty \frac{1}{ 1-  q^{i+\frac12}}\prod_{j=0}^\infty \frac{1}{ 1-q^{j+\frac12}}=
 \frac{1}{(q^\frac12,q)_\infty^2},
 \\
  I_R&=&\prod_{i=0}^\infty \frac{1}{ 1+(-1)^iq^{i+\frac12}}\prod_{j=0}^\infty \frac{1}{ 1+(-1)^jq^{j+\frac12}}
   \\
 &=&
   \frac{1}{ (-q^\frac12,-q)_\infty ^2},
   \qquad
 \eeqn
which  are   again consistent with the limits of the    (unflavored) Macdonald index below
\beqn
\cI_M&=& 
\frac{1}{((qT)^{\frac12}  ;q)_\infty^2}.
\eeqn
Note that in this case, the  Macdonald index above contains fractional powers of $q,T$, but our specific prescription \eqref{IRind} for  $\cI_R$ is still unambiguous. 

\subsection{$(A_1,A_{2m})$}  

Next, we  consider  various kinds of Argyres-Douglas theories. The first class is the $(A_1,A_{2m})$ AD theories with $m\in \bZ_{>0}$.
The VOAs of these theories   have  only one strong generator, the stress tensor. This type of VOA gives rise to the 
Virasoro minimal models  \cite{Belavin:1984vu}, which  are labelled by a pair of   integers $ (\sfp,\sfq)$, subject to the condition  $\sfq>\sfp\ge 2,\gcd(\sfp,\sfq)=1$, and have central charges
\be
c_{\sfp,\sfq}=1-\frac{6(\sfp-\sfq)^2}{\sfp\sfq}.
\ee
The VOA of  $(A_1,A_{2m})$ AD is just    given by the $(2,2m+3)$ Virasoro minimal model.

The automorphism map $\phi$ acts as $\phi(T)=T$, and the inner product is $\EB{T,T}= c_{2,2m+3}/2 <0$, which is consistent with \eqref{posO}, because $T$, coming from the  multiplet $\hat \cC_{0(0,0)} $, has $h=2, R=1$ and $r=0$.

The Macdonald index of  $(A_1,A_{2m})$ AD theory can be computed using various methods and admits several equivalent descriptions. One simple form is given by   \cite{Foda:2019guo} 
\be \label{A1A2mMac}
\cI^{(A_1,A_{2m})}_{ {M}}  = \sum_{N_1 \geq \cdots \geq N_m \geq 0}^{\infty} \frac{q^{N_1^2 + \cdots + N_m^2  } (qT)^{N_1 + \cdots + N_m}}{(q)_{N_1 - N_2} \cdots (q)_{N_{m-1} - N_m}(q)_{N_m}}  .
\ee

Using our proposal, we can  now compute the limits of the Macdonald index.  
For $m=1$, our proposal \eqref{Idr} gives
\be
 I^{(A_1,A_{2 })}_{ {R}}  =1 - q^2 + q^3 - q^4 + q^5 + q^8 - q^9 + 2q^{10} - 2q^{11} + \cdots,
\ee
which is indeed consistent with \eqref{A1A2mMac}.
We have checked the agreement up to order $q^{12}$ for all $m<10$.

It is   believed that the \((\sfp,\sfq)\) minimal model with \(\sfp,\sfq \ge 3\) cannot be realized as the VOA of any 4d SCFT. Nevertheless, our procedure for computing \(I_R\) is still applicable. In particular, we can compute \(I_R\) for the unitary Virasoro minimal models with \(\sfq=\sfp+1\). It turns out that \(I_R(q)=I_S(-q)\) in this case. For other non-unitary minimal models, one can perform   similar computations. For example, for \((3,5)\), we find
\be
I^{(3,5)}_R=1 - q^2 + q^3 - 2q^4 + 2q^5 - 2q^6 + 2q^7 - 2q^8 + q^9 - q^{11} +\cdots.
\ee
 This differs from the conventional vacuum character of the \((3,5)\) minimal model and suggests that our modified character may have broader applications beyond the SCFT/VOA context.

\subsection{$(A_{N-1},A_{k-1  }) $}   
The previous class of AD theories belongs to a  more general family of  Argyres-Douglas theories, known as $(A_{N-1},A_{k-1  })$ theories, which are labelled by a pair of coprime integers $\gcd(k,N)=1$.
Without loss of generality we can restrict to the case $ k> N$.  The VOA is proposed in \cite{Cordova:2015nma} to  be given by    the $\cW_N$ minimal model  \cite{Fateev:1987vh,Fateev:1987zh}.

In general, the $\cW_N(\sfp,\sfq)$ minimal models are indexed by two coprime integers $\sfq>\sfp\ge N$ with central charge \cite{Bouwknegt:1992wg,Fateev:1987vh,Fateev:1987zh}
\be
c^{(N)}_{ \sfp, \sfq}=(N-1)\Big( 1-\frac{N(N +1)(\sfp-\sfq)^2}{\sfp\sfq} \Big). 
\ee
The VOA of the $(A_{N-1},A_{k-1  })$ AD  theory is given by $\cW_N(N,k+N)$ minimal model  \cite{Cordova:2015nma}. 

The $\cW_N$ VOA has strong generators $W_2\equiv T, W_3, \cdots, W_N$ with $h_{W_i}=i$. The OPE is known but complicated.

For the \(\cW_3\) VOA, the OPE is relatively simple and is presented in Appendix~\ref{opeVOA}. The algebra has two strong generators, \(W_2 \equiv T\) and \(W_3 \equiv W\). We can determine the anti-linear automorphism by imposing the defining conditions, together with \eqref{posO} and the appropriate \(R\)-charge selection rules \footnote{From the OPE in Appendix~\ref{opeVOA}, it follows immediately that \(\phi(W)=sW\) with \(s=\pm1\). We then compute \(\EB{W,W}=sc/3\), and use \eqref{posO} to deduce that \(s=-(-1)^{h_W+R_W-r_W}=(-1)^{R_W}\), where we used \(c<0\) and \(r_W=0\). The latter follows from the relation \(r_W=-r_{\phi(W)}\). To determine \(R_W\), we first note that \(R_W<h_W=3\), since the relation \(R=h\) holds only for Higgs branch operators, which are absent in this theory. Thus the possible values are \(R_W=0,1,2\). We then use the fact that, for a singular term in the OPE, the \(R\)-charges satisfy \(R_{\{AB\}_n}\le R_A+R_B-1\) for \(n>0\). Applying this to the term \(\Lambda\subset W\times W\), where \(\Lambda\) is the quasi-primary operator built from \(T\) with \(h_\Lambda=4\) and \(R_\Lambda=2\), uniquely fixes \(R_W=2\), and hence \(s=1\). This also suggests that \(W\) arises from the multiplet \(\hat{\cC}_{1(0,0)}\).}. The final result is   \(\phi(W)=W\).

There is no known simple general formula for the Macdonald index in this case, but the Schur index has the following  form 
\be
\cI_S=\mathrm{PE}\Big[\frac{ q^2 (1 - q^{k - 1}) (1 - q^{N - 1}) }{ (1 - q)^2 (1 - q^{k + N} ) }
\Big].
\ee
Our procedure   \eqref{Ids} reproduces the Schur index. Moreover,  we can use \eqref{Idr}  to compute $I_R$:
\beqn
I_R^{(A_2,A_3)}&=& 1 - q^2 + q^4 - q^5 + q^7 - q^8 + 2q^9 +\cdots  , \qquad  \\
I_R^{(A_2,A_7)}&=& 1 - q^2 + q^4 +  q^8 +q^9+\cdots  ,  
\eeqn
which are consistent with the Macdonald index computed with the method  in \cite{Song:2016yfd}.  We have checked for all $k<10$.

We also studied the $\cW_4$ case. In the simplest case, our proposal gives
\be
I_R^{(A_3,A_4)}=  1 - q^2 + q^5 - q^6 + q^8 + q^9+\cdots. 
\ee

\subsection{$(A_1,D_{2n+1})$}
Next, we consider   $(A_1,D_{2n+1})$ AD theories. This class of theories has flavor symmetry $\fks\fku(2)$, and the corresponding VOA is given by the Kac-Moody algebra $\widehat{\fks\fku(2)}_k$ at level $k =-\frac{4n}{2n+1}$ \cite{Cordova:2015nma}.  The OPE is given by
\be
J^a(z) J^b(0) \sim \frac{\frac{k}{ 2} \delta^{ab}}{z^2}+\frac{i \epsilon_{abc} J^c}{z}, \qquad a=1,2,3~,
\ee
where  $\epsilon$ is the   antisymmetric symbol with $\epsilon_{123}=1$. The Kac-Moody algebra has the   anti-linear automorphism  $\phi(J^a)=-  J^a $, where the minus sign is related to the presence of the imaginary factor in the OPE. The inner product $\EB{J^a,J^b}=-\frac{k}{ 2} \delta^{ab}>0$. This is again consistent with \eqref{posO}, because $J^a$, coming from the  multiplet $\hat \cB_{1} $, have $h= R=1$ and $r=0$.

We can then compute the index using our proposal. For $n=1$, we get
\beqn
I_S &\! =&\!  1 + 3q + 9q^2 + 19q^3 + 42q^4 + 81q^5 + 155q^6+\cdots, \qquad
\\ 
I_R &\!=&\! 1 + 3q + q^2 + 3q^3 + 10q^4 + 3q^5 + 11q^6+\cdots,
\eeqn
which indeed agrees with the  limits of Macdonald index given  in Appendix~\ref{macId}. We have explicitly  checked   all the cases $n<10$.

\subsection{$\cT_{(3,2)}$}
Our final example is given by the so-called $\cT_{(3,2)}$   theories, which are constructed as follows: one first takes three copies of   $ (A_1,D_3) $  theories, and then   gauges their diagonal  flavor   $\fks\fku(2)$ symmetry. It turns out that this   type of gauging is conformal and  the resulting    theory $\cT_{(3,2)}$ has the nice property that $a_{4d}=c_{4d}$. 

It was proposed in \cite{Buican:2020moo} that the VOA of the $\cT_{(3,2)}$ theory is given by the $\cA(6)$ chiral algebra  \cite{Feigin:2007sp,feigin2008characters}, which has three strong generators, $T, \Psi, \tilde\Psi$, with weight 2, 4, 4, respectively. The first one is the stress tensor and the last two are fermionic. The full OPE can be found in \cite{Jiang:2024baj}; see also  Appendix~\ref{opeVOA}. 
One can show that the anti-linear automorphism is
\be
\phi(T)=T, \quad \phi(\Psi)=t\tilde\Psi, \quad\phi(\tilde\Psi)=\frac{1}{t}\Psi , \qquad t\in \mathbb R.
\ee
 The inner products are then  $\EB{\Psi,\Psi}=6t, \EB{\tilde\Psi,\tilde\Psi}= -6/t$. Just like the case of free vector, without loss of generality, we can choose $t=1$.   
  The condition \eqref{posO} implies $R_\Psi-r_\Psi\in 2\bZ$,  $R_{\tilde\Psi}-r_{\tilde\Psi}=R_{ \Psi}+r_{ \Psi}\in 2\bZ+1$, which further  suggests that $R_\Psi,r_\Psi \in 2\bZ+\frac12$. Actually it can be shown that $R_\Psi=\frac52, r_\Psi=1/2$  
\footnote{Since \(0\le R\le h\), the condition $R_\Psi,r_\Psi \in 2\bZ+\frac12$ implies that \(R_\Psi=\frac12\) or \(\frac52\). The possibility \(R_\Psi=\frac12\) can be excluded, since it would require \(\Psi\) to belong to a multiplet of type \(\cD_{0(0,j_2)}\) or \(\bar{\cD}_{R(j_1,0)}\), and such multiplets occur only in free theories, with \(j_i=0\). We therefore conclude that \(R_\Psi=\frac52\).
Moreover, the structure of the Macdonald index at order \(q^4\) takes the form \(q^4(T+T^2-T^2-T^3)\), where the term \(q^4T^2\) corresponds to the quasi-primary operator \(\Lambda\). The remaining negative contributions must then come from \(\Psi\) and \(\tilde\Psi\), which implies \(R_\Psi-r_\Psi=2\) and \(R_\Psi+r_\Psi=3\). Hence \(R_\Psi=\frac52\) and \(r_\Psi=\frac12\).
Examining the multiplet structure of Schur operators, one then finds that \(\Psi\in \hat{\cC}_{\frac32(0,\frac12)}\) and \(\tilde\Psi\in \hat{\cC}_{\frac32(\frac12,0)}\).
}.


Using our procedure, we find  
\beqn
I_S&=&  1 +  {q^2} + q^3 + 2q^6 + q^8 + q^{11}+\cdots
\\
I_R&=&1 - q^2+ q^3 + q^8 -q^{11}+\cdots
\eeqn
which indeed agree  with the special limits of the Macdonald index   computed in Appendix~\ref{macId}. 

\section{Defect generalization} \label{defect}

We now further generalize the proposal to the case with surface defects, which correspond to non-vacuum modules of the VOA  \cite{Cordova:2017mhb}.

We will illustrate with Virasoro minimal model  corresponding to $(A_1,A_{2m})$  AD theory. To this aim, we  now   switch to the language of states and mode  algebra. Given a highest weight $\ket{\sfh}$, satisfying $L_0\ket{\sfh}=\sfh\ket{\sfh},L_n\ket{\sfh}=0 $ for $n>0$, we can generate the Verma module
\be
\ket{\mathbf n}
= L_{-n_s}\cdots L_{-n_2}L_{-n_1}\ket{\sfh}, \qquad  
\ee
where $0<  n_1\le n_2\le \cdots \le n_s$.  $L_n$'s are the modes of the stress tensor and satisfy the Virasoro algebra. 
Using operator-state correspondence, the  anti-linear automorphism defined before can be translated into the following action on the Virasoro modes   
$
\phi(L_n)=L_{-n}  
$,
which is conventionally   written as $(L_n)^\dagger =L_{-n}$ and indeed leaves the Virasoro algebra invariant.
The inner product   is  now given by
\be\label{mnh}
\EB{\mathbf m , \mathbf n}_\sfh=(-1)^N\bra{\sfh} L_{m_1} \cdots   L_{ m_t}L_{-n_s}\cdots  L_{-n_1}\ket{\sfh},
\ee
which is  zero unless $N\equiv \sum_i m_i =\sum_j n_j$.   Note that the extra sign factor comes from the   operator-state correspondence.  
This defines a   Hermitian form.  Our claim is that we can use this new product and repeat our previous procedure to    calculate the  defect generalization of indices in  \eqref{IdS} and \eqref{IRind}. 

We   consider a special class of surface operators in $(A_1,A_{2m})$  theories which  correspond to the module   of the $(2,2m+3)$  Virasoro minimal model with $\sfh_s=  \frac{(1-s) (2 m-s+2)}{4 m+6}$, $s=1, \cdots, m+1$.  Here $s=1$ corresponds to vacuum module without surface operator insertion. 
The Macdonald defect index in this case has been computed 
\cite{Foda:2019guo}
 \be\label{defMd}
\cI^{(A_1,A_{2m})}_{ {M},s} =\!\! \!\!\!\! \sum_{N_1 \geq \cdots \geq N_m \geq 0}^{\infty} \frac{q^{N_1^2 + \cdots + N_m^2  +N_s + \cdots + N_m}
T^{N_1 + \cdots + N_m}}
{(q)_{N_1 - N_2} \cdots (q)_{N_{m-1} - N_m}(q)_{N_m}}.
\ee
 Using our prescription around \eqref{Ids}\eqref{Idr} with generalized inner product \eqref{mnh},   we find that the $T=1$ and $T=-1,q\to -q $ limit can be recovered exactly. For example, in the simplest case of  $m=1,s=2,h_s=-\frac15$, our formula  \eqref{Ids}  and \eqref{Idr} give
 \beqn
 I_S &=&   1 + q + q^2 + q^3 + 2q^4 + 2q^5 + 3q^6 + 3q^7+O(q)^8,
 \qquad\quad
 \\
 I_{R}&=& 1 - q^2 + q^3 - q^4 + q^5+O(q)^8,
 \eeqn
which agree  with the special limits of  \eqref{defMd}. We have verified this for all $m\le 10, s\le 11$ up to order $O(q)^8$.  The agreement of these results strongly suggests that the notion of graded unitarity still holds in the current case with defects.

\section{Conclusion}\label{conl}

In this letter, we have proposed a universal and intrinsic method for computing a distinguished limit of the Macdonald index directly from the VOA. This limit is different from the Schur limit and naturally gives rise to a new type of modified character intrinsic to the VOA. Our construction requires no additional assumptions and applies to all unitary 4d SCFTs.

Several important questions remain open. A first direction is to develop methods for computing this quantity to all orders and, ideally, to obtain closed-form expressions for broad classes of VOAs.
 A second direction is to investigate its number-theoretic properties. Since the modified character is again a series in \(q\) or \(q^{\frac12}\), it is natural to ask whether it satisfies any modular properties. For ordinary characters, modularity plays a central role in the study of module structures and in the classification of VOAs \cite{Beem:2017ooy}. It would be very interesting to study  whether similar structures extend to our modified character, even beyond the setting of the SCFT/VOA correspondence.

Our construction   is based on an anti-linear automorphism of the VOA satisfying a number of nontrivial properties. From the viewpoint of the SCFT/VOA correspondence, this automorphism is related to CPT conjugation in the parent 4d theory. In all examples we have studied, such an anti-linear automorphism exists and  can be determined by   various conditions. It would be desirable to establish this statement in greater generality, extend the analysis to broader classes of VOAs, and classify such automorphisms systematically.

Finally, it is important to apply our procedure to a wider class of SCFTs, including other Argyres--Douglas theories, \(\mathcal N=4\) super-Yang--Mills theory, and \(\cN=3\) theories. In particular, the case with defects, which is related to non-vacuum modules, deserves further investigation. To place the defect generalization of our proposal on a firmer foundation, one should formulate the notion of graded unitarity in the presence of defects and justify the corresponding extension of our procedure.

\bigskip
\noindent
\emph{Acknowledgement.}
This work was supported by the startup
grant at SIMIS and the Shanghai Pujiang Program (No. 25PJA128).

\appendix
\section{Unitarity constraint on the inner product}\label{posOpf}
Our goal in this appendix is to prove the condition \eqref{posO} using the notion of graded unitarity established in \cite{ArabiArdehali:2025fad}.

For this purpose, we need another anti-linear automorphism \(\rho\), introduced in \cite{ArabiArdehali:2025fad}. This map satisfies the defining properties
\[
\rho(\mathbf{1})=\mathbf{1}, \qquad \rho(T)=T,
\]
\[
\rho(uA+vB)=u^*\rho(A)+v^*\rho(B),
\]
and
\be\label{rhoAB}
\rho\big(\{AB\}_n\big)=(-1)^{f_Af_B}\{\rho(A)\rho(B)\}_n.
\ee
Moreover, one has \cite{ArabiArdehali:2025fad}
\be\label{rhosq}
\rho^2=(-1)^{2R}=(-1)^{2(r-h)},
\ee
where the second equality follows from
$
R-r+h=2(R+j_1)\in \bZ,
$
which is a consequence of \eqref{Schcond}.

The anti-linear map \(\rho\) allows us to define the sesquilinear form
\be
\EV{A,B}=\{\rho(A)B\}_{h_A+h_B},
\ee
where \(\{AB\}_n\) denotes the coefficient of the order-\(n\) pole in the OPE \(A(z)B(0)\). Since \(\{\rho(A)B\}_{h_A+h_B}\) is the most singular term in the OPE, it is either zero or proportional to the identity operator; in the definition above, the identity operator is stripped off. Using the general property of the OPE \eqref{ABope}, one readily finds
\be\label{evab2}
\EV{A,B}=(-1)^{h_A+h_B}(-1)^{f_Af_B}\{B\rho(A)\}_{h_A+h_B}.
\ee
This defines a sesquilinear form, namely
\[
\EV{uA,vB}=u^*v\,\EV{A,B}.
\]

Under complex conjugation, we have
\beqn
\EV{A,B}^*
&=&
\rho\!\left(\{\rho(A)B\}_{h_A+h_B}\right)
\nonumber\\
&=&
(-1)^{f_Af_B}\{\rho^2(A)\rho(B)\}_{h_A+h_B}
\nonumber\\
&=&
\rho_A^2\,(-1)^{f_Af_B}\{A\rho(B)\}_{h_A+h_B}
\nonumber\\
&=&
\rho_A^2\,(-1)^{h_A+h_B}\EV{B,A},
\eeqn
where we used \eqref{rhoAB}, the fact that \(f_{\rho(A)}=f_A\), and \eqref{evab2}. In what follows, we will only consider the case \(h_A=h_B\) and \(f_A=f_B\); otherwise, for quasi-primary operators \(A\) and \(B\), the inner product vanishes. In this case,
\be
\rho_A^2(-1)^{h_A+h_B}
=
(-1)^{2(r_A-h_A)}(-1)^{2h_A}
=
(-1)^{2r_A}
\ee
where we used \eqref{rhosq}.  Note that 
for Schur operators, spin-statistics implies \( (-1)^{2r}=(-1)^f \).
Consequently,
\be\label{EVconj}
\EV{A,B}^*=(-1)^{2r_A}\EV{B,A}.
\ee
One may further verify that
\[
\EV{A,B}^{**}=\EV{A,B}.
\]
Thus \(\EV{-,-}\) is Hermitian for bosonic operators and anti-Hermitian for fermionic operators.

To obtain a Hermitian form uniformly, we introduce a new anti-linear automorphism
\be
\phi=\rho\circ i^{ 2r}=i^{ 2r}\circ\rho.
\ee
More explicitly,
\beqn
\phi(A)
&=&
\rho\big(i^{ 2r_A}A\big)
\\
&=&
(-i)^{ 2r_A}\rho(A)
=
i^{-2r_A}\rho(A)
\\
&=&
i^{  2r_{\rho(A)}}\rho(A)
=
(i^{ 2r}\circ\rho)(A),
\eeqn
where we used the fact that \(\rho\) is anti-linear and reverses the \(U(1)_r\) charge. It is then straightforward to verify that
\be
\phi^2=(-1)^{2h}.
\ee
Indeed,
\beqn
\phi^2(A)
&=&
\phi\big(\phi(A)\big)
\\
&=&
i^{ 2r}\circ\rho\big(i^{-2r_A}\rho(A)\big)
\\
&=&
(-i)^{-2r_A}\,i^{2r}\circ\rho\big(\rho(A)\big)
\\
&=&
(-i)^{-2r_A}\,i^{ 2r}\big(\rho^2(A)\big)
\\
&=&
(-i)^{-2r_A}i^{ 2r_A}(-1)^{2R_A}A
\\
&=&
(-1)^{2(R_A+r_A)}A
\\
&=&
(-1)^{2h_A}A,
\eeqn
where in the last step we used
$
h_A-R_A-r_A=2j_{1,A}\in\mathbb Z,
$
which follows from \eqref{Schcond}.

We can now define a new inner product
\be
\EB{A,B}=\{\phi(A)B\}_{h_A+h_B},
\ee
which is related to the previous one by
\beqn
\EB{A,B}
&=&
\{i^{-2r_A}\rho(A)B\}_{h_A+h_B}
\\
&=&
i^{-2r_A}\{\rho(A)B\}_{h_A+h_B}
\\
&=&
i^{-2r_A}\EV{A,B}.
\label{EBEVrel}
\eeqn
Under complex conjugation,
\beqn
\EB{A,B}^*
&=&
(-i)^{-2r_A}\EV{A,B}^*
\\
&=&
(-i)^{-2r_A}(-1)^{-2r_A}\EV{B,A}
\\
&=&
i^{-2r_A}\EV{B,A}
\\
&=&
\EB{B,A},
\eeqn
where we used \eqref{EVconj} and \(r_B=r_A\). Therefore, \(\EB{-,-}\) is Hermitian, and this is the form used in the main text.

We are now ready to prove \eqref{posO}. In \cite{ArabiArdehali:2025fad}, it was shown that for the two-point function in the VOA,
\be
\EV{\cO(z) (\rho\circ\cO)(0)}=\frac{\kappa_\cO}{z^{2h_\cO}},
\ee
the coefficient satisfies
\be\label{kappaOs}
\kappa_\cO \propto i^{2h_\cO-2R_\cO},
\ee
where the proportionality factor is positive.  Note that this condition holds both for quasi-primary operators and their descendants:
 if \eqref{kappaOs} is valid for $\cO$, it should also hold for its descendants, following from   $\kappa_{\cO'}=-2 ( 1 + 2 h_\cO) h_\cO\kappa_\cO$,  and $h_{\cO'}=h_\cO+1$, $R_{\cO'}=R_\cO$. 

Comparing with our notation in \eqref{evab2}, we obtain
\beqn
\EV{\cO,\cO}
&=&
(-1)^{2h_\cO}(-1)^{f_\cO^2}\{\cO\rho(\cO)\}_{2h_\cO}
\\
&=&
(-1)^{2h_\cO}(-1)^{f_\cO}\kappa_\cO,
\eeqn
where we used the fact that \(f^2-f=0\mod 2\) for \(f\in\bZ\). Using the relation \eqref{EBEVrel}, we then find
\beqn
\EB{\cO,\cO}
&=&
i^{-2r_\cO}\EV{\cO,\cO}
\\
&=&
i^{-2r_\cO}(-1)^{2h_\cO}(-1)^{f_\cO}\kappa_\cO
\\
&=&
i^{-2r_\cO}i^{-4h_\cO}i^{4r_\cO}\kappa_\cO
\\
&\propto&
i^{-4h_\cO}i^{2r_\cO}i^{2h_\cO-2R_\cO}
\\
&\propto&
i^{-2h_\cO+2r_\cO-2R_\cO}
\\
&\propto&
(-1)^{h_\cO+R_\cO-r_\cO},
\eeqn
where we used \eqref{kappaOs}. This proves \eqref{posO}.

  \section{Some explicit OPEs   }\label{opeVOA}
  
  The  $\cW_3$ VOA  has two strong  bosonic generators, whose OPEs are given by 
  \begin{align}
T(z)T(0)
&\sim
\frac{c/2}{z^4}
+\frac{2\,T(0)}{z^2}
+\frac{\partial T(0)}{z},
\\[6pt]
T(z)W(0)
&\sim
\frac{3\,W(0)}{z^2}
+\frac{\partial W(0)}{z},
\\[6pt]
W(z)W(0)
&\sim
\frac{c/3}{z^6}
+\frac{2\,T(0)}{z^4}
+\frac{\partial T(0)}{z^3}
\nonumber\\
&\quad
+\frac{1}{z^2}
\left(
\frac{32}{22+5c}\,\Lambda(0)
+\frac{3}{10}\,\partial^2 T(0)
\right)
\nonumber\\
&\quad 
+\frac{1}{z}
\left(
\frac{16}{22+5c}\,\partial \Lambda(0)
+\frac{1}{15}\,\partial^3 T(0)
\right),
\end{align}
 where 
\be
\Lambda =T^2-\frac{3}{10}\,\partial^2 T ,
\ee
is   a quasi-primary operator with dimension 4.

    The  $\cA(6)$  VOA  has three strong    generators, one bosonic $T$ and two fermionic $\Psi, \tilde \Psi$. The  corresponding OPEs are given by 
  
\begin{align}
T(z)T(0) &\sim -\frac{12}{z^4}+\frac{2T}{z^2}+\frac{T'}{z},\\[4pt]
T(z)\Psi(0) &\sim \frac{4\Psi}{z^2}+\frac{\Psi'}{z},\\[4pt]
T(z)\tilde{\Psi}(0) &\sim \frac{4\tilde{\Psi}}{z^2}+\frac{\tilde{\Psi}'}{z},\\[4pt]
\Psi(z)\tilde{\Psi}(0)
&\sim
-\frac{6}{z^8}
+\frac{2T}{z^6}
+\frac{T'}{z^5}
+\frac{3\left(T''-T^2\right)}{7z^4}
\nonumber\\
&\qquad 
+\frac{2T^{(3)}-9T'T}{21z^3}
\nonumber\\
&\qquad 
+\frac{-48(T')^2-84T''T+36T^3+7T^{(4)}}{420z^2}
\nonumber\\
&\qquad
+\frac{60\left(-5T''T'+6T'T^2-2T^{(3)}T\right)+7T^{(5)}}{2800z}.
\end{align}

\section{Some formulae for Macdonald index}\label{macId}

The Macdonald index of the $ (A_1,D_{2n+1})$ Argyres-Douglas theories is given by  \cite{Song:2015wta}
\beqn
&&
I_M^{(A_1,D_{2n+1})}
=
\frac{1}{(qT z^{\pm 2,0};q)_\infty}
\sum_{\lambda= 0}^\infty
(-1)^\lambda
q^{\lambda(\lambda+1)\left(n+\frac12\right)}
T^{\lambda(n+1)}
\nonumber \\&&\qquad\qquad\qquad 
\times
\frac{(q^{\lambda+1};q)_\lambda\,(T^2 q^{2\lambda+2};q)_\infty}
{( T q^{\lambda+1};q)_\lambda\,(T q^{2\lambda+2};q)_\infty}
P_{2\lambda}(z),
\eeqn 
where  
\[
  (x;q)_n=\prod_{i=0}^{n-1}(1-x q^i), \qquad
   (q)_n\equiv (q;q)_n,
\]
and\[
P_\lambda(z)
=
\sum_{i=0}^{\lambda}
\frac{(t;q)_i}{(q;q)_i}
\frac{(t;q)_{\lambda-i}}{(q;q)_{\lambda-i}}
z^{2i-\lambda}.
\]
In the above equations, $z$ is the flavor fugacity for $\fks\fku (2)$ flavor symmetry. 

For $n=1$, the unflavored Macdonald index reads
\beqn
\cI_M^{(A_1,D_{3})}\!\! 
&=&\!
1 + 3Tq + (4T + 5T^2)q^2 + (4T + 8T^2 + 7T^3)q^3
\nonumber\\&&
+ (4T + 17T^2 + 12T^3 + 9T^4)q^4
+ (4T + 23T^2 
\nonumber\\&&
+ 27T^3 + 16T^4 + 11T^5)q^5
+ (4T + 33T^2 + 48T^3 
\nonumber\\&&
+ 37T^4 + 20T^5 + 13T^6)q^6
+ O(q)^7.
\eeqn

The $T\to 1$ limit of $\cI_M^{(A_1,D_{2n+1})}$ gives the Schur index, which takes the following simple form
\[
\cI_S^{(A_1,D_{2n+1})}
=
\mathrm{PE}\!\left[
\frac{q-q^{2n+1}}{(1-q)(1-q^{2n+1})}\,\Big( 1+z^2+\frac{1}{z^2}\Big)
\right]~.
\]

We next compute the Macdonald index of the $ \cT_{(3,2)}$ Argyres-Douglas theories, which are obtained from the diagonal gauging of   three copies of $D_3(SU(2))=(A_1,D_3)=(A_1,A_3)$ theories. 
This type of gauging procedure allows us to compute the Macdonald index  as follows:
\be
\cI_M ^{ \cT_{(3,2)}}=\frac12 \oint \frac{dz}{2\pi iz}  (1-z^2)(1-z^{-2}) \Big( \cI_M^{(A_1,D_3)}  \Big) ^3  \cI_M^ \text{vec} ,
\ee
where 
\be
 \cI_M^\text{vec}(q,T,z)=\mathrm{PE}\Big[\frac{-q(1+T)}{1-q}\Big( 1+z^2+\frac{1}{z^2}\Big) \Big],
\ee
is the  Macdonald index of free vector multiplet. After performing the integral, we find that
\[ 
\begin{aligned}\label{Md32indx}
\cI_M^{ \cT_{(3,2)}} &=1 + q^2 T + q^3 T + q^4 (T - T^3) + q^5 (T - T^3) + q^6 (T + T^2) \\
&\quad + q^7 (T + T^2 - T^3 - T^4) + q^8 (T + 2T^2 - T^3 - T^4) \\
&\quad + q^9 (T + 2T^2 - T^3 - 2T^4) + q^{10} (T + 3T^2 - T^3 - 3T^4) \\
&\quad + q^{11} (T + 3T^2 - T^3 - 3T^4 + T^6)+\cdots~.
\end{aligned}
\]

  \bigskip
\hfill 

 \bibliography{ref}

\end{document}